\documentclass[aps,pra,floatfix,twocolumn,superscriptaddress,showpacs,10pt]{revtex4-1}
\usepackage{amssymb,amsmath,color,mciteplus,graphicx,subfigure}
\usepackage{calc,epsfig,epstopdf,color,mciteplus,bm,mathrsfs}
\usepackage{times}
\usepackage{float}
\usepackage{lipsum}

\begin{document}

\title{Universal Robust Geometric Quantum Control via Geometric Trajectory Correction}

\author{Tao Chen} 
\affiliation{Key Laboratory of Atomic and Subatomic Structure and Quantum Control (Ministry of Education), School of Physics, South China Normal University, Guangzhou 510006, China}
\affiliation{Guangdong Provincial Key Laboratory of Quantum Engineering and Quantum Materials, Guangdong-Hong Kong \\ Joint Laboratory of Quantum Matter, Frontier Research Institute for Physics, South China Normal University, Guangzhou 510006, China}

\author{Jia-Qi Hu}
\affiliation{Key Laboratory of Atomic and Subatomic Structure and Quantum Control (Ministry of Education), School of Physics, South China Normal University, Guangzhou 510006, China}

\author{Chengxian Zhang}\email{cxzhang@gxu.edu.cn}
\affiliation{School of Physical Science and Technology, Guangxi University, Nanning 530004, China}

\author{Zheng-Yuan Xue}\email{zyxue83@163.com}
\affiliation{Key Laboratory of Atomic and Subatomic Structure and Quantum Control (Ministry of Education), School of Physics, South China Normal University, Guangzhou 510006, China}
\affiliation{Guangdong Provincial Key Laboratory of Quantum Engineering and Quantum Materials, Guangdong-Hong Kong \\ Joint Laboratory of Quantum Matter, Frontier Research Institute for Physics, South China Normal University, Guangzhou 510006, China}


\begin{abstract}
  Universal robust quantum control is essential for performing complex quantum algorithms and efficient quantum error correction protocols. Geometric phase, as a key element with intrinsic fault-tolerant feature, can be well integrated into quantum control processes to enhance control robustness. However, the current geometric quantum control is still controversial in \emph{robust universality}, which leads to the unsatisfactory result that cannot sufficiently enhance the robustness of arbitrary type of geometric gate. In this study, we find that the finite choice on geometric evolution trajectory is one of the main roots that constrain the control robustness of previous geometric schemes, as it is unable to optionally avoid some trajectory segments that are seriously affected by systematic errors. In view of this, we here propose a new scheme for universal robust geometric control based on geometric trajectory correction, where enough available evolution parameters are introduced to ensure that the effective correction against systematic errors can be executed. From the results of our numerical simulation, arbitrary type of geometric gate implemented by using the corrected geometric trajectory has absolute robustness advantages over conventional quantum one. In addition, we also verify the feasibility of the high-fidelity physical implementation of our scheme in superconducting quantum circuit, and finally discuss in detail the potential researches based on our scheme. Therefore, our theoretical work is expected to offer an attractive avenue for realizing practical fault-tolerant quantum computation in existing experimental platforms.
\end{abstract}

\maketitle

\section{Introduction}

Quantum computation utilizes unique effects such as quantum superposition and quantum entanglement for information encoding and processing \cite{QC}, and thus can hold the permit to solve some complex problems more efficiently than classical computation. However, current experimental manipulation on available quantum hardware is still vulnerable to environment noise-induced decoherence and inaccurate control. Thus, one of the keys to realizing large-scale quantum computation at this stage is the research on robust quantum control \cite{FTQC}, aiming at applying practical fault-tolerant methods to handle different types of errors in quantum systems. In this way, the universal robust quantum control with low error sensitivity can be utilized to effectively enhance the utilization of circuit resources in performing quantum algorithms and quantum error correction protocols.

Geometric quantum computation, as an attractive gate construction strategy, is characterized by integrating the geometric phases \cite{Abelian-adiabatic, nonAbelian-adiabatic, Abelian-nonadiabatic, nonAbelian-nonadiabatic} with intrinsic fault-tolerant features into the quantum control processes to realize universal robust geometric quantum gate. To alleviate the long-time consumption loss induced by strict adiabatic conditions in the previous adiabatic geometric quantum computation \cite{AHQC, NMRAGQC, ionAHQC}, nonadiabatic geometric quantum computation (NGQC) \cite{NGphaseG, NGQC, NHQC, NHQC-DFS, NGQC-NHQC} based on nonadiabatic geometric phases \cite{Abelian-nonadiabatic, nonAbelian-nonadiabatic} has been put forward to shorten the gate time and simplify the required control. The highlight of fast and simple control of NGQC has thus inspired many experimental follow-up on various quantum systems, including trapped ions \cite{Geo-ion1, Geo-ion2}, NMR \cite{Geo-NMR1, Geo-NMR2, Geo-NMR3}, superconducting circuits \cite{Geo-SQ1, Geo-SQ2, Geo-SQ3, Geo-SQ4, Geo-SQ5, Geo-SQ6, Geo-SQ7}, NV centers \cite{Geo-NV1, Geo-NV2, Geo-NV3, Geo-NV4, Geo-NV5}, etc.

However, the current NGQC still has shortcomings in the \emph{universality} of its robust geometric control \cite{RobustQ1}, which results in the fact that the implemented geometric gates are too sensitive to systematic error and inevitably causes the loss of gate fidelity. Recently, a widely-adopted method for solving the above issue is to integrate optimal control techniques \cite{NGQC+DD, NGQC+Com, NHQC+, NGQC+TOC, NGQC+GP, Geo+Op1} into the geometric control process, but the resulting enhancement of robustness is only limited to a few special types of geometric gate and/or only targeted at a certain type of error. Apart from that, unlike the single-qubit case, the coupling interaction between two qubits in most quantum systems is usually unable to be tuned in an arbitrary and time-dependent manner, thus making it difficult to apply mechanically the optimal control techniques in the construction of two-qubit geometric gates to enhance their robustness. Therefore, it is still necessary to return to the source to determine how to realize universal robust geometric control in a better way without additional optimal control techniques.

In this study, we first determine that the finite choice on geometric evolution trajectory is one of the main roots that constrain the control robustness of previous geometric schemes \cite{NGQC-Path11, NGQC-Path12, NGQC-Path21, NGQC-Path22}, because it is unable to optionally avoid some trajectory segments that are seriously affected by systematic errors. Therefore, we here propose a new scheme for universal robust geometric control based on geometric trajectory correction, where enough available evolution parameters are introduced to ensure that the effective correction against systematic errors can be executed. In this way, arbitrary type of geometric gate (including two-qubit entanglement gates) implemented by using our corrected geometric trajectory can show absolute robustness advantages over conventional quantum gates, which is also verified by our numerical simulation. Furthermore, we then take the superconducting quantum circuit as an example to illustrate the feasibility of the high-fidelity physical implementation of our scheme. Finally, we discuss in detail the potential researches that can be further carried out based on our scheme.

\section{Conventional quantum gates}

We begin by introducing a gate-construction strategy that is favored in current experiments \cite{Nature14,DRAGExp}. For a two-level quantum structure of $\{|0\rangle, |1\rangle\}$ driven by an external microwave field, its Hamiltonian in the rotating frame can be written as
\begin{eqnarray} \label{genH}
\mathcal{H}(t)=\frac {1} {2}
\left(
\begin{array}{cccc}
 -\Delta(t)             & \Omega(t) e^{-i\phi(t)} \\
 \Omega(t) e^{i\phi(t)} & \Delta(t)
\end{array}
\right),
\end{eqnarray}
where $\Omega(t)$ and $\phi(t)$ are the time-dependent driving amplitude and phase of the microwave field, and $\Delta(t)$ is the frequency detuning between the microwave field and the resonance of the two-level system. Conventionally, to implement a universal set of quantum gates, the case of the resonance transition $|0\rangle\!\leftrightarrow\!|1\rangle$ and a constant phase is considered, that are $\Delta(t)=0$ and $\phi(t)=\phi_c$. The resulting evolution operator is given by
\begin{eqnarray} \label{ConU}
U_c(\theta_c,\phi_c)&=&e^{-i\int^{\tau}_0 \mathcal{H}(t) \textrm{d}t}  \notag\\
&=&\left(
\begin{array}{cccc}
 \cos\frac{\theta_c}{2}             & -i \sin\frac{\theta_c}{2} e^{-i\phi_c} \\
 -i \sin\frac{\theta_c}{2} e^{i\phi_c} & \cos\frac{\theta_c}{2}
\end{array}
\right),
\end{eqnarray}
where $\theta_c=\int^{\tau}_0 \Omega(t) \textrm{d}t$ represents an arbitrary rotation angle. So by setting $\phi_c=0$ and $\pi/2$, the rotation operations around $X-$ and $Y-$axis denoted as $X^c_{\theta_c}$ and $Y^c_{\theta_c}$ are realized, respectively. In addition, the rotation operations $X^c_{-\theta_c}$ and $Y^c_{-\theta_c}$ for negative angles need to reset $\phi_c=\pi$ and $-\pi/2$. In this way, arbitrary type of conventional quantum gate can be obtained by a combination of $X^c_{\pm\theta_c}$ and/or $Y^c_{\pm\theta_c}$, such as identity gate $I^c=X^c_{2\pi}$ and Hadamard gate $H^c=Y^c_{-\frac{\pi}{2}}X^c_{\pi}$.

\begin{figure}[tbp]
  \centering
  \includegraphics[width=0.9\linewidth]{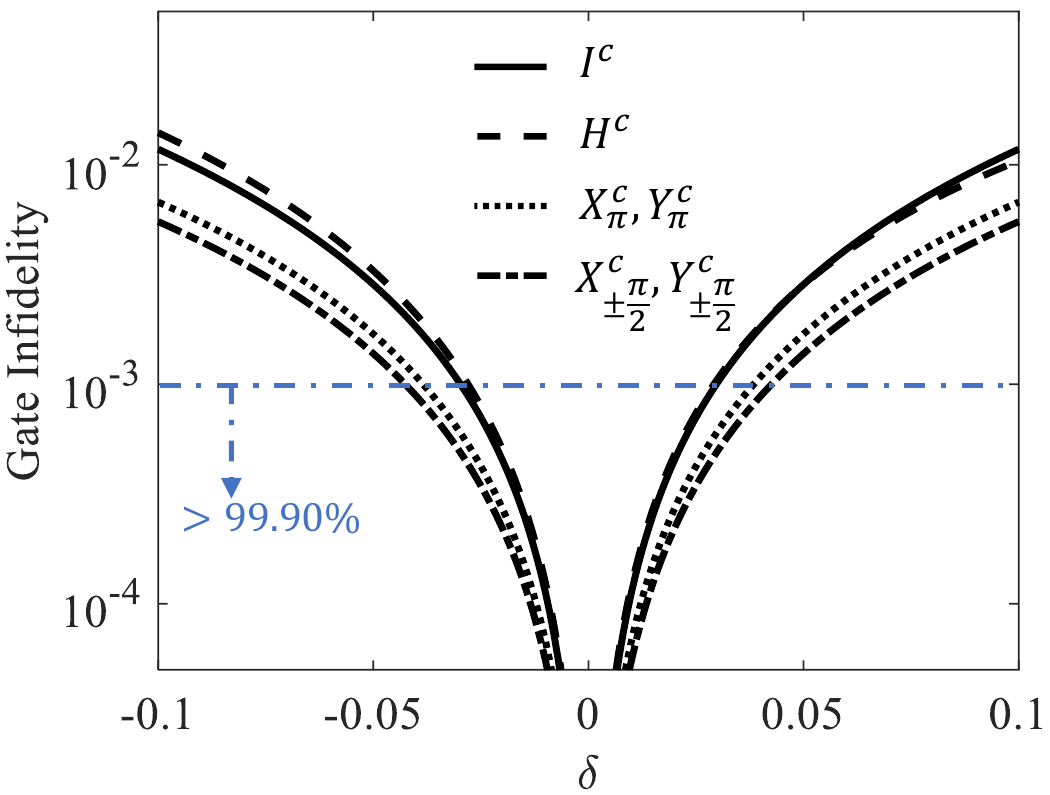}
  \caption{Sensitivity of conventional $I^c$, $H^c$, $X^c_{\pi}$, $Y^c_{\pi}$, $X^c_{\pm\frac{\pi}{2}}$ and $Y^c_{\pm\frac{\pi}{2}}$ gates to systematic error, in which we take the detuning error in the form of $\Delta(t)\rightarrow\Delta(t)+\delta\Omega_m$ as an example (without decoherence), and consider a simple time-dependent pulse $\Omega(t)=\Omega_m \sin\frac{\pi t}{\tau}$. Below the blue dashed line corresponds to the regions where the single-qubit gate fidelity is higher than $99.90\%$. The calculation formula of gate error sensitivity is $1-F_U$ with $F_U\!=\!\textrm{Tr}(U^{\dagger}U_{e})/\textrm{Tr}(U^{\dagger}U)$ being the definition of gate fidelity, where $U$ and $U_{e}$ represent the desired gate and the gate affected by the systematic error, respectively.} 
  \label{GatePDyn}
\end{figure}

In current experiments \cite{Nature14,DRAGExp} based on the above conventional gate-construction strategy, the average single-qubit gate fidelity has reached a remarkable result of more than $99.90\%$. However, from the numerical simulation results as shown in Fig. \ref{GatePDyn}, we can find that it is difficult to maintain high-fidelity performance for all the affected conventional quantum gates, when further considering the systematic error commonly existed in experiments. But quantum gates with low error sensitivity are crucial for performing complex quantum algorithms and efficient quantum error correction protocols. Thus, how to reduce the sensitivity of quantum gates to systematic error while maintaining high fidelity is the key to realize large-scale fault-tolerant quantum computation, and this is also the core problem we aim to solve here.

\section{General Method of Gate Construction}

\begin{figure*}[tbp]
  \centering
  \includegraphics[width=0.95\linewidth]{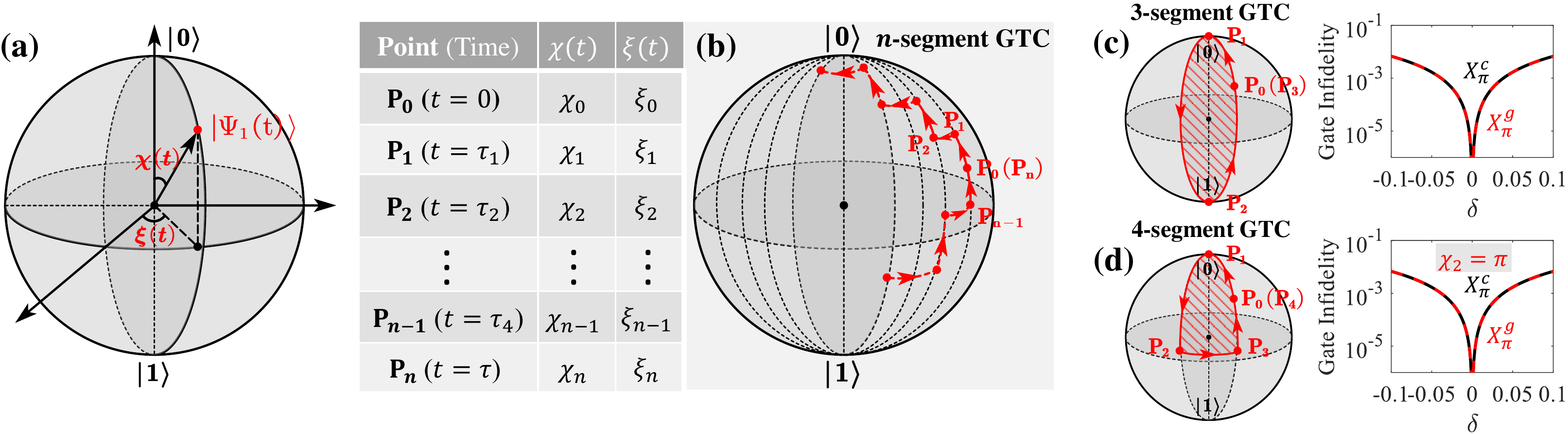}
  \caption{(a) The position coordinates of state vector $|\Psi_1(t)\rangle$ at a certain moment on the Bloch sphere. (b) The evolution details of our general multi-segment geometric trajectory, in which $\tau_i$ represents the time point between the initial time $t=0$ and the final time $t=\tau$. The evolution details involved in executing (c) $3$-segment geometric trajectory correction (abbreviated as GTC in the figure) and (d) $4$-segment geometric trajectory correction, and their corresponding correction results in which the black line (red line) is the sensitivity of the conventional $X^c_{\pi}$ gate (geometric $X^g_{\pi}$ gate realized based on the corrected geometric trajectory) to systematic error.}
  \label{Path}
\end{figure*}

Based on the Hamiltonian $\mathcal{H}(t)$ of the driven two-level system in Eq. (\ref{genH}), we will work to propose the general method of quantum gate construction, in which the flexible design of the time-dependent Hamiltonian parameters $\Omega(t)$, $\phi(t)$ and $\Delta(t)$ is utilized to implement different gate-construction strategies. That way, there are more options for determining optimal and robust quantum computation scheme.

In general, the evolution operator governed by Hamiltonian $\mathcal{H}(t)$ can be solved by the equation $U(t)=\mathcal{T}e^{-i\int \mathcal{H}(t) \textrm{d}t}$, in which $\mathcal{T}$ is the time-ordering operator.
However, when the complex time-dependent shapes of the Hamiltonian parameters $\Omega(t)$, $\phi(t)$ and $\Delta(t)$ are involved during optimal control, an accurate result of evolution operator is hard to solve directly.
Thus, to get a simple and general solution of the evolution operator, we next consider two orthogonal evolution states
\begin{subequations} \label{2state}
\begin{align}
|\Psi_1(t)\rangle&=e^{if_1(t)}[\cos\frac{\chi(t)}{2}|0\rangle+\sin\frac{\chi(t)}{2} e^{i\xi(t)}|1\rangle], \\
|\Psi_2(t)\rangle&=e^{if_2(t)}[\sin\frac{\chi(t)}{2}e^{i\xi(t)}|0\rangle -\cos\frac{\chi(t)}{2} |1\rangle],
\end{align}
\end{subequations}
where $\chi(t)$ and $\xi(t)$ are the spherical coordinates of the state vector on the Bloch sphere as shown in Fig. \ref{Path}(a), and $f_{k}(t)$ is the overall phase accumulated, with $f_{k}(0)=0$ at time $t=0$.
As we all know, the evolution process of the state $|\Psi_k(t)\rangle$ is governed by the Hamiltonian $\mathcal{H}(t)$, and they are connected by the Schr\"{o}dinger equation $i(\partial/\partial t) |\Psi_k(t)\rangle =\mathcal{H}(t)|\Psi_k(t)\rangle$. So we can solve that the correspondences between state-evolution parameters $\{\chi(t),\xi(t)\}$ and Hamiltonian parameters are
\begin{subequations} \label{paralimt}
\begin{align}
\dot{\chi}(t)&=\Omega(t)\sin[\phi(t)-\xi(t)], \\
\dot{\xi}(t)&=-\Delta(t)-\Omega(t)\cot\chi(t)\cos[\phi(t)-\xi(t)].
\end{align}
\end{subequations}
In addition, the overall phase accumulated during the evolution period $\tau$ is written as
\begin{eqnarray} \label{phaselimt}
\gamma=f_1(\tau)=-f_2(\tau)
=\int^{\tau}_0\! \frac{\dot{\xi}(t)\left[1-\cos\chi(t)\right]\!+\!\Delta(t)}{2\cos\chi(t)} \textrm{d}t. \quad \ \
\end{eqnarray}
Therefore, by comparing the changes of two orthogonal evolution states at the initial moment $t=0$ and the final moment $t=\tau$, we can ultimately obtain the general solution of evolution operator (governed by Hamiltonian $\mathcal{H}(t)$) as
\begin{eqnarray} \label{generalU}
U(\tau)=\sum_{k=1,2}|\Psi_k(\tau)\rangle\langle \Psi_k(0)|.
\end{eqnarray}
It is clear that the time-dependent shapes of the Hamiltonian parameters $\Omega(t)$, $\phi(t)$ and $\Delta(t)$ directly determines the evolution details of state vector $|\Psi_k(t)\rangle$ characterized by $\xi(t)$, $\chi(t)$ and $\gamma$. Meanwhile, according to Eq. (\ref{generalU}), we only need to fix the boundary values of the evolution parameters $\xi(t)$ and $\chi(t)$ to realize arbitrary type of quantum gate. Thus, it is expected that the goal-directed setting for the shapes of Hamiltonian parameters can be done to optimize the evolution process of state vector, so as to realize universal robust quantum control.

Using the above general method of quantum gate construction, the conventional gate in Eq. (\ref{ConU}) can also be realized by engineering the evolution trajectory of state $|\Psi_k(t)\rangle$ to evolve along the longitude line, that is, setting $\dot{\xi}(t)=0$. However, as discussed and verified numerically before, it is difficult for conventional quantum gates to maintain high-fidelity performance under the influence of systematic error. In the following, we will focus on developing the effective universal robust control that can reduce the error sensitivity of quantum gates while maintaining high gate fidelity.

Taking the state vector $|\Psi_1(t)\rangle$ as an example, the dynamic properties of its evolution process are determined by the accumulated overall phase $\gamma$ in Eq. (\ref{phaselimt}), including the dynamical phase
\begin{eqnarray} \label{gammaD}
\begin{split}
\gamma_d&=-\int^\tau_0\langle \Psi_1(t)|\mathcal{H}(t)|\Psi_1(t)\rangle \textrm{d}t \\
&=\frac{1}{2}\int^{\tau}_0 \frac{\dot{\xi}(t)\sin^2\chi(t)+\Delta(t)}{\cos\chi(t)} \textrm{d}t,
\end{split}
\end{eqnarray}
and the geometric phase
\begin{eqnarray} \label{gammaG}
\gamma_g=\gamma-\gamma_d=-\frac {1} {2}\int^\tau_0 \dot{\xi}(t)\left[1-\cos\chi(t)\right] \textrm{d}t,
\end{eqnarray}
where the geometric nature \cite{Abelian-nonadiabatic} of $\gamma_g$ comes from the fact that it is given by half of the solid angle enclosed by the evolution trajectory. Based on it, we first aim to achieve geometric robust protection of the evolution process by strictly satisfying $\Delta(t)=-\dot{\xi}(t)\sin^2\chi(t)$ to make $\gamma_d=0$. And then, we continue to correct the geometric evolution trajectory by flexibly engineering the time-dependent shapes of $\chi(t)$ and $\xi(t)$ (correspond to the Hamiltonian parameters $\Omega(t)$, $\phi(t)$ and $\Delta(t)$), so as to further enhance the robustness of geometric control against systematic error.

\section{Geometric Trajectory Correction}

\begin{figure*}[tbp]
  \centering
  \includegraphics[width=0.95\linewidth]{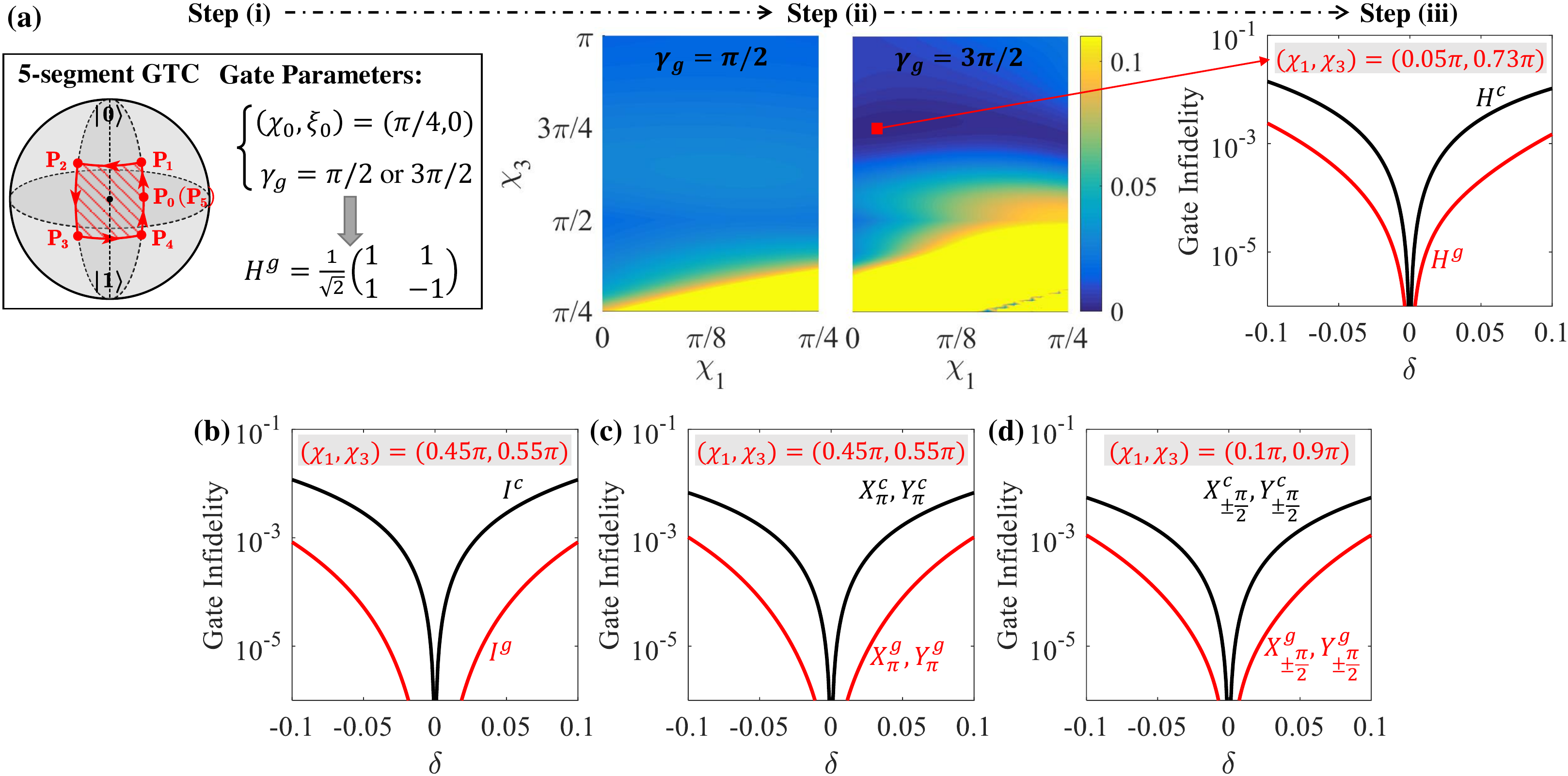}
  \caption{(a) Detailed steps to realize universal robust geometric gates using the $5$-segment geometric trajectory correction. We take the geometric Hadamard gate $H^g$ as a typical example: \textbf{(i)} firstly, determine its corresponding gate parameters $(\chi_0, \xi_0)=(\pi/4, 0)$, and $\gamma_g=\pi/2$ or $3\pi/2$; \textbf{(ii)} fix the optional ranges as $\chi_1\in[0,\pi/4]$ and $\chi_3\in[\pi/4,\pi]$ according to the gate parameter $\chi_0=\pi/4$, and then analyse the error sensitivity of different $5$-segment geometric trajectories determined by optional $\chi_1$ and $\chi_3$, in which the error quantity considered is $\delta=0.1$; \textbf{(iii)} finally, utilize the corrected $5$-segment geometric trajectory, that determined by the optimal regions of $\chi_1=0.05\pi$ and $\chi_3=0.73\pi$ with low error sensitivity, to implement a robust geometric $H^g$ gate. In the same way, other robust geometric (b) $I^g$, (c) $X^g_{\pi}$, $Y^g_{\pi}$, (d) $X^g_{\pm\frac{\pi}{2}}$, $Y^g_{\pm\frac{\pi}{2}}$ gates (red line) can also be demonstrated a stronger suppression on systematic error than the conventional counterparts (black line).}
  \label{GateP1}
\end{figure*}

Next, we will propose a universal robust geometric control scheme based on geometric trajectory correction, in which the evolution process characterized by $\chi(t)$ and $\xi(t)$ is shown in Fig. \ref{Path}(b). The general evolution trajectory given here starts from point $\textbf{P}_\textbf{0}[\chi_0,\xi_0]$, passes through a total of $n$ segments of longitude and latitude lines, and then loops back to the initial position $\textbf{P}_\textbf{0}[\chi_0,\xi_0]$ at the final time $\tau$, where evolution parameter $\chi(t)$ ($\xi(t)$) is a time-dependent variable on the longitude (latitude) line of the trajectory, and $\dot{\chi}(t)=0$ ($\dot{\xi}(t)=0$) on the latitude (longitude) line of the trajectory. In this way, the accumulated pure geometric phase, after removing the dynamical phase $\gamma_d$, is given by
$\gamma_g=-\sum_{n} (\xi_{n+1}-\xi_n)\left[1-\cos\chi_n\right]/2$. The resulting evolution operator can be expanded as
\begin{eqnarray} \label{GeoUall}
&&\!\!\!\!U_g(\chi_0,\xi_0,\gamma_g)= \\
&&\left(
\begin{array}{cccc}
\cos\gamma_g+i\sin\gamma_g\cos\chi_0 & i \sin\gamma_g\sin\chi_0 e^{-i\xi_0} \\
i \sin\gamma_g\sin\chi_0 e^{i\xi_0}  & \cos\gamma_g-i\sin\gamma_g\cos\chi_0
\end{array}
\right), \notag
\end{eqnarray}
where we only need to set the initial point of the cyclic trajectory, namely $\chi_0$ and $\xi_0$, and geometric phase $\gamma_g$ to implement arbitrary geometric gate. Such as, setting $(\chi_0,\xi_0)=(\pi/2,\pi)$ and $(\pi/2,-\pi/2)$ with the same $\gamma_g=\theta_g/2$ (or $\theta_g/2+\pi$), the geometric rotation operations around $X-$ and $Y-$axis, denoted as $X^g_{\theta_g}$ and $Y^g_{\theta_g}$, can be obtained, respectively. And for the geometric rotation operations $X^g_{-\theta_g}$ and $Y^g_{-\theta_g}$ with negative angles, it needs to reset $\xi_0=0$ and $\pi/2$. Thus, the geometric identity gate is $I^g=X^g_{2\pi}$, and the geometric Hadamard gate $H^g$ can be directly determined by setting $(\chi_0,\xi_0)=(\pi/4,0)$ and $\gamma_g=\pi/2$ (or $3\pi/2$). It can be seen that after assigning $\chi_0$ and $\xi_0$ according to the targeted gate type, we can also further correct the geometric trajectory by re-selecting the remaining evolution parameters $\{\chi_1, \chi_2,..., \chi_{n-1}\}$ or $\{\xi_1, \xi_2,..., \xi_{n-1}\}$, thereby utilizing the corrected $n$-segment geometric trajectory to achieve more universal and robust geometric control, which can effectively resist the influence of systematic error on gate construction.

For constructing a universal set of quantum gates, the special $1$-segment geometric trajectory correction and $2$-segment geometric trajectory correction are obviously invalid, so these two cases are ignored here. Due to the lack of available evolution parameters, the $3$-segment geometric trajectory shown in Fig. \ref{Path}(c) cannot be further corrected, and the geometric gate implemented based on it does not show a stronger suppression on systematic error than conventional quantum gate (taking $X_{\pi}$ gate as an example). Furthermore, as shown in Fig. \ref{Path}(d), we perform the $4$-segment geometric trajectory correction by optimizing the available parameter $\chi_2$ (or $\xi_2$). However, it can be found that the optimal result based on $4$-segment geometric trajectory correction is obtained precisely at a special $\chi_2=\pi$, in which the $4$-segment geometric trajectory correction recovers to the fixed $3$-segment geometric trajectory and holds the same result. Therefore, the $3$-segment and $4$-segment geometric trajectory corrections still play a limited role in suppressing systematic error, as they are unable to utilize enough available evolution parameters to optionally avoid some trajectory segments that are seriously affected by systematic error. Note that the $3$-segment and $4$-segment geometric trajectory corrections here correspond to the previous single-loop \cite{NGQC-Path11, NGQC-Path12} and short-path NGQC schemes \cite{NGQC-Path21, NGQC-Path22}, respectively. So, it is necessary to continue the multi-segment ($n>4$) geometric trajectory correction to ensure that there are enough available evolution parameters to execute the effective correction, so as to realize universal robust control and show the absolute advantages in suppressing systematic error compared with conventional quantum gates.

\section{Multi-segment Geometric Trajectory Correction}

In this section, we continue to perform $5$-segment geometric trajectory correction and verify its effectiveness for achieving universal robust geometric control. According to the evolution details of the $5$-segment trajectory in Fig. \ref{GateP1}(a), we can obtain the specific settings of Hamiltonian parameters through the known parameter correspondence in Eq. (\ref{paralimt}). For example, in the first time segment $t\in[0,\tau_1]$, the trajectory evolves from initial point $\textbf{P}_\textbf{0}(\chi_0,\xi_0)$ to point $\textbf{P}_\textbf{1}(\chi_1,\xi_1)$ along the longitude line, where the corresponding evolution parameter satisfies $\dot{\xi}(t)=0$ and $\xi(t)=\xi_0=\xi_1$, so resulting in the simplified parameter correspondence as
\begin{subequations} \label{paralimtsim1}
\begin{align}
\chi_1-\chi_0&=\int^{\tau_1}_0\Omega(t)\sin[\phi(t)-\xi_0]\textrm{d}t, \\
\Delta(t)&=-\Omega(t)\cot\chi(t)\cos[\phi(t)-\xi_0].
\end{align}
\end{subequations}
In addition, we also need to set $\phi(t)-\xi_0=-\pi/2$ in the case of $\chi_0>\chi_1$ to ensure that the pulse area $\int^{\tau_1}_{0}\Omega(t) \textrm{d}t$ is positive and minimal. By substituting $\phi(t)=\xi_0-\pi/2$ into Eq. (\ref{paralimtsim1}), the setting of Hamiltonian parameters in the first time segment $t\in[0,\tau_1]$ can be obtained. Then, in the second time segment $t\in[\tau_1,\tau_2]$, the trajectory continues to evolve from $\textbf{P}_\textbf{1}(\chi_1,\xi_1)$ to $\textbf{P}_\textbf{2}(\chi_2,\xi_2)$ along the latitude line, where the corresponding evolution parameter satisfies $\dot{\chi}(t)=0$ and $\chi(t)=\chi_1=\chi_2$, so resulting in the simplified parameter correspondence as
\begin{eqnarray} \label{paralimt21}
\phi(t)=\pi+\xi(t), \quad \dot{\xi}(t)=-\Delta(t)+\Omega(t)\cot\chi_1,
\end{eqnarray}
or
\begin{eqnarray} \label{paralimt22}
\phi(t)=\xi(t), \quad \dot{\xi}(t)=-\Delta(t)-\Omega(t)\cot\chi_1.
\end{eqnarray}
Note that, the choice of these two results depends on whether it can make the pulse area $\int^{\tau_2}_{\tau_1}\Omega(t) \textrm{d}t$ strictly positive after determining the evolution parameters $\chi_1$ and $\chi_3$. Here, taking the result in Eq. (\ref{paralimt21}) as an example, we can further substitute $\Delta(t)\!=\!-\dot{\xi}(t)\sin^2\chi_1$ (geometric condition) into it to solve the setting of Hamiltonian parameters in the second time segment $t\in[\tau_1,\tau_2]$. Based on the same way, the setting of Hamiltonian parameters for the remaining time segments can also be obtained, and not presented in detail here. Therefore, the all settings of Hamiltonian parameters corresponding to $5$-segment geometric trajectory are given by
\begin{subequations} \label{OmegaPhi}
\begin{align}  \label{OmegaPhia}
\!\!\!\!\!&t\in[0,\tau_1]:
\int^{\tau_1}_0\Omega(t)\textrm{d}t=\chi_0-\chi_1, \ \
\phi(t)=\xi_0-\frac{\pi}{2},
\\
\!\!\!\!\!\begin{split} \label{OmegaPhib}
t\in[\tau_1,\tau_2]:
\int^{\tau_2}_{\tau_1}\Omega(t)\textrm{d}t=(\xi_2-\xi_0)\sin\chi_1 \cos\chi_1,
\\
\!\!\!\phi(t)=\xi_0+\pi+\frac{\int \Omega(t)\textrm{d}t}{\sin\chi_1 \cos\chi_1},
\end{split}
\\
\!\!\!\!\!&t\in[\tau_2,\tau_3]:
\int^{\tau_3}_{\tau_2}\Omega(t)\textrm{d}t=\chi_3-\chi_1,
\phi(t)=\xi_2+\frac{\pi}{2},
\\
\!\!\!\!\!\begin{split}
t\in[\tau_3,\tau_4]:
\int^{\tau_4}_{\tau_3}\Omega(t)\textrm{d}t=(\xi_0-\xi_2)\sin\chi_3 \cos\chi_3,
\\
\!\phi(t)=\xi_2+\pi+\frac{\int \Omega(t)\textrm{d}t}{\sin\chi_3 \cos\chi_3},
\end{split}
\\
\!\!\!\!\!&t\in[\tau_4,\tau]:
\int^{\tau}_{\tau_4}\Omega(t)\textrm{d}t=\chi_3-\chi_0,\ \
\phi(t)=\xi_0-\frac{\pi}{2},
\end{align}
\end{subequations}
with $\Delta(t)=0$, $-\Omega(t)\tan\chi_1$, $0$, $-\Omega(t)\tan\chi_3$ and $0$, respectively. The final geometric evolution operator is also shown in Eq. (\ref{GeoUall}), where $\gamma_g=(\xi_2-\xi_0)[\cos\chi_1-\cos\chi_3]/2$ is a pure geometric phase accumulated in the whole $5$-segment geometric evolution trajectory.

It can be seen that, under the $5$-segment geometric trajectory, the implementation of arbitrary geometric gate requires only fixing parameters $\chi_0$, $\xi_0$ and $\gamma_g$. Therefore, the free optimization of the available evolution parameters $\chi_1$ and $\chi_3$, in the range of $[0,\chi_0]$ and $[\chi_0,\pi]$, can develop the flexibility in the design of Hamiltonian parameters $\{\Omega(t), \Delta(t), \phi(t)\}$, and provide us more options to effectively correct the $5$-segment geometric trajectory against the influence of systematic error. The detailed steps to realize robust geometric gates using the $5$-segment geometric trajectory correction are as follows:

\textbf{(i)} firstly, determine gate parameters $\{\chi_0, \xi_0, \gamma_g\}$ required to obtain the target geometric gate by Eq. (\ref{GeoUall});

\textbf{(ii)} secondly, fix the optional ranges of $\chi_1$ and $\chi_3$ according to the gate parameter $\chi_0$, that are $[0,\chi_0]$ and $[\chi_0,\pi]$, and then analyse the error sensitivity of different $5$-segment geometric trajectories determined by optional $\chi_1$ and $\chi_3$;

\textbf{(iii)} correct the geometric trajectory through continuous optimization of $\chi_1$ and $\chi_3$, and finally determine the one with low error sensitivity, which is used to implement a robust geometric gate superior to conventional quantum gate \cite{Nature14,DRAGExp}.

As shown in Fig. \ref{GateP1}(a), an example of the robust geometric Hadamard gate $H^g$ implemented based on the steps \textbf{(i)}-\textbf{(iii)} of $5$-segment geometric trajectory correction is given to illustrate. In the same way, arbitrary type of robust geometric gate can also be implemented, and their optimal correction results are shown in Figs. \ref{GateP1}(b)-(d). Therefore, we can find that our $5$-segment geometric trajectory correction executed by optimizing the available parameters $\chi_1$ and $\chi_3$ can effectively achieve universal robust geometric control, thus greatly reducing the error sensitivity of arbitrary type of geometric gate, not just a few special quantum gate types. As an extension, it is worth noting that as the number $n$ of trajectory segments increases, our multi-segment geometric trajectory correction will be able to effectively suppress more error sources simultaneously, or demonstrate better suppression effect for certain errors.
\section{Feasibility of physical implementation}

\begin{figure}[tbp]
  \centering
  \includegraphics[width=0.95\linewidth]{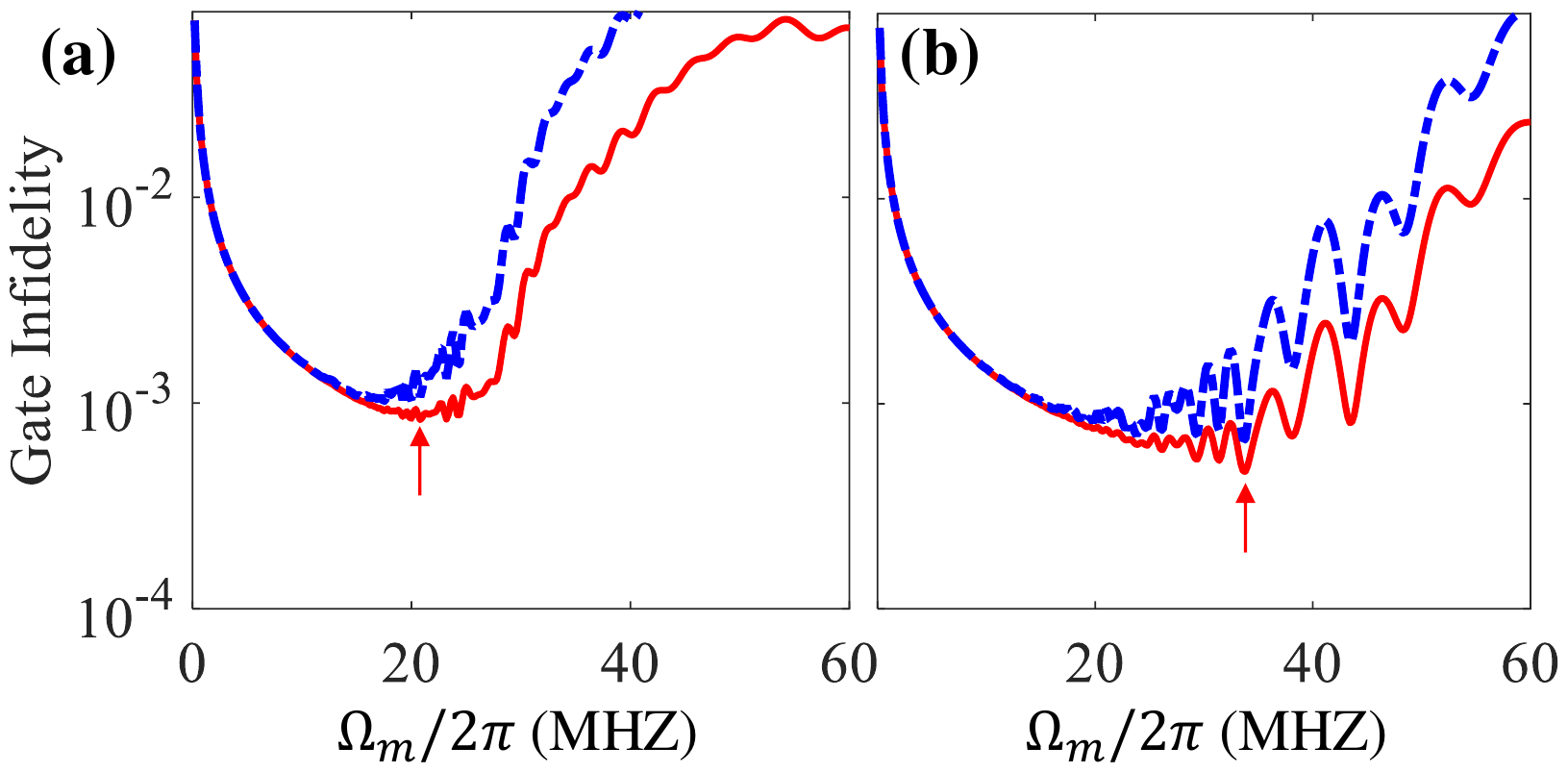}
  \caption{Gate infidelity of the implemented geometric (a) $H^g$ and (b) $X^g_{\frac{\pi}{2}}$ gates as function of tunable parameter $\Omega_m$, with DRAG correction (red solid line) and without DRAG correction (blue dashed line), where the red arrow points to the point with the highest gate fidelities of (a) 99.91\% and (b) 99.93\%.}
  \label{PhyIm1}
\end{figure}

In the above discussion, the implementation of our geometric trajectory correction scheme under the Hamiltonian $\mathcal{H}(t)$ in Eq. (\ref{genH}) only needs to involve a simple two-level structure, which is easily obtained by addressing in the realistic physical systems, including superconducting circuit \cite{SQ-twoL, Geo-SQ5}, NV center \cite{adiabaticGeo-NV, NV-twoL}, trapped ion \cite{ion-twoL}, etc. We next consider a scalable and flexibly controllable superconducting quantum system to verify the feasibility of our geometric scheme in physical implementation. First, for a single superconducting transmon-type qubit driven by a microwave field, its whole Hamiltonian form contains, in addition to the effective Hamiltonian $\mathcal{H}(t)$ in Eq. (\ref{genH}) used to realize our targeted geometric control, the unwanted leakage Hamiltonian as follows:
\begin{eqnarray} \label{leakH}
\mathcal{H_{\textrm{leak}}}(t)&=& \frac{1}{2}\sum^{\infty}_{n=2}\left[(2n-1)\Delta(t)-n(n-1)\alpha_1\right] |n\rangle\langle n| \notag\\
&+& \frac{1}{2}\sum^{\infty}_{n=2}\sqrt{n}\left[\Omega(t)e^{-i\phi(t)}|n-1\rangle\langle n| +\mathrm{H.c.}\right] \quad
\end{eqnarray}
with $\alpha_1$ being the intrinsic anharmonicity of transmon, which is caused by the spurious transitions from the qubit subspace $\{|0\rangle, |1\rangle\}$ to the higher levels and will greatly limit gate fidelity. Besides, the decoherence effect induced by the inevitable coupling of quantum system to its surrounding environment is also an important consideration in physical implementation. We include these two effects into the quantum dynamics simulated by the Lindblad master equation \cite{DRAG1}, in which we set a conservative parameter region with the transmon anharmonicity $\alpha_1=2\pi\times320$ MHz and the relaxation time ($T_1$) and pure dephasing time ($T_{\phi}$) about $50$ $\mu s$ according to state-of-the-art technologies of superconducting qubits \cite{Decay1,Decay2}. Take the geometric $H^g$ and $X^g_{\frac{\pi}{2}}$ gates implemented based on $5$-segment geometric trajectory correction (corresponding to optimal parameters $(\chi_1, \chi_3)=(0.05\pi,0.73\pi)$ and $(0.1\pi,0.9\pi)$) as examples and their simulated results are shown in Figs. \ref{PhyIm1}(a) and \ref{PhyIm1}(b),
we can find that limited by the effects of leakage error (proportional to $\Omega_m$) and decoherence (inverse proportional to $\Omega_m$), their resulting gate fidelities (blue dashed line) can not reach a high level. Therefore, we here introduce the DRAG (Derivative Removal by Adiabatic Gate) correction procedure \cite{DRAG1,DRAG2} to combat the effect of leakage error therein to gate fidelity and to minimize gate-operation time. The final correction results (red solid line) in Figs. \ref{PhyIm1}(a) and \ref{PhyIm1}(b) verify that our geometric trajectory correction scheme can also be well compatible with the optimal control techniques to solve the practical problems existed in the superconducting implementation, and achieve the high fidelities of 99.91\% and 99.93\% at $\Omega_m\approx2\pi\times21$ MHz and $\Omega_m\approx2\pi\times34$ MHz for geometric $H^g$ and $X^g_{\frac{\pi}{2}}$ gates, respectively.

\begin{figure*}[tbp]
  \centering
  \includegraphics[width=0.95\linewidth]{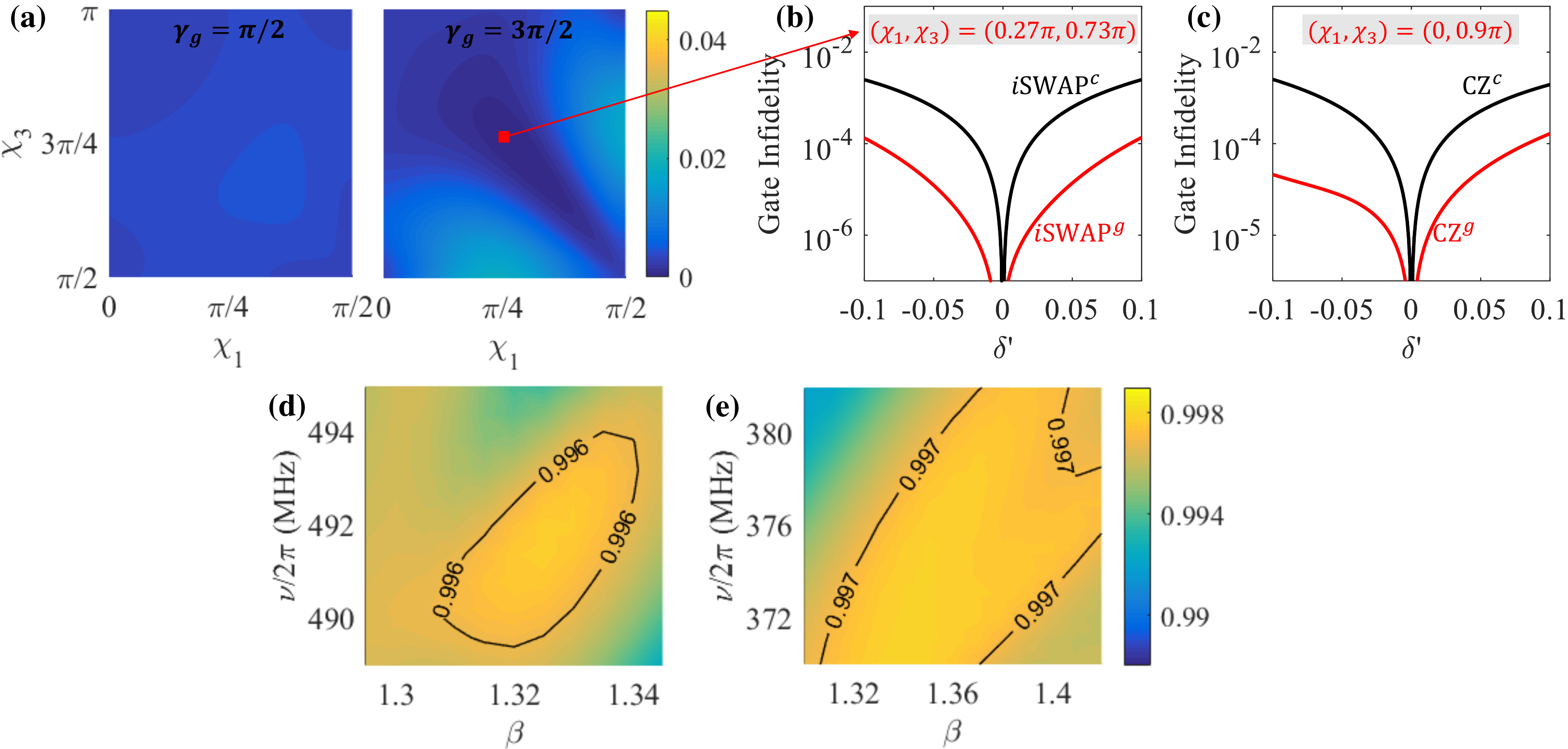}
  \caption{(a) The error sensitivity of our geometric $i\textrm{SWAP}^g$ gate under different 5-segment geometric trajectories determined by $\chi_1$ and $\chi_3$ (the optional ranges as $\chi_1\in[0,\pi/2]$ and $\chi_3\in[\pi/2,\pi]$, respectively), where the error quantity considered is $\delta'=0.1$. Under the optimal regions of $(\chi_1,\chi_3)=(0.27\pi,0.73\pi)$ and $(0,0.9\pi)$, the fault-tolerant advantages of our robust geometric (b) $i\textrm{SWAP}^g$ and (c) $\textrm{CZ}^g$ gates (red line) compared to the conventional $i\textrm{SWAP}^c$ and $\textrm{CZ}^c$ gates (black line). Gate fidelity of geometric (d) $i\textrm{SWAP}^g$ and (e) $\textrm{CZ}^g$ gates in superconducting implementation as functions of tunable parameters $\nu$ and $\beta$, in which the overall influences of the neglected higher-order oscillating terms and the system decoherence are taken into account.}
  \label{GateP2}
\end{figure*}

Generally, our geometric trajectory correction scheme can also be applied to the construction of superconducting two-qubit entanglement gates, thus which combine the implemented arbitrary single-qubit gates to accomplish universal fault-tolerant quantum computation. Here we adopt a direct capacitive coupling between two transmons $Q_1$ and $Q_2$ (with anharmonicities $\alpha_1$ and $\alpha_2$), in which the qubit frequency of $Q_1$ is driven by an additional parametric modulation \cite{TunCoupleExp1,TunCoupleExp2,TunCoupleExp3} and is of the form $\omega_1(t)=\omega_1+\varepsilon\sin(\nu t+\varphi)$, where $\varepsilon$, $\nu$, and $\varphi$ are the modulation frequency, amplitude, and phase, respectively. Indexing the excited states of coupled qubits as $|Q_1, Q_2\rangle$, the obtained interaction Hamiltonian can be written as
\begin{eqnarray} \label{EqHITwo}
\!\!\!\mathcal{H}_I(t)&=& g \sum^{+\infty}_{m=-\infty}i^m J_m(\beta)
\left\{ |01\rangle\langle 10|e^{i\Delta_1 t}e^{im(\nu t+\varphi)} \right.
\notag \\
&&\left.+\sqrt{2}|02\rangle\langle 11|e^{i(\Delta_1-\alpha_2) t}e^{im(\nu t+\varphi)}\right.
\notag \\
&&  \left.+\sqrt{2}|11\rangle\langle 20|e^{i(\Delta_1+\alpha_1 t)}e^{im(\nu t+\varphi)}\right\} +\mathrm{H.c.},
\end{eqnarray}
where $g$ and $\Delta_1=\omega_2-\omega_1$ is the coupling strength and the frequency difference between the two transmons $Q_1$ and $Q_2$ respectively, and $J_m(\beta)=J_m(\varepsilon/\nu)$ is the $m$th order Bessel function of the first kind. Therefore, the tunable coupling interaction within the single-excitation subspace $\{|01\rangle, |10\rangle\}$ or the two-excitation subspace $\{|02\rangle, |11\rangle\}$ can be determined by different settings of the modulation frequency $\nu$ and the ratio $\beta$, that is
\begin{subequations} \label{set2qubit1}
\begin{align}
\!\!\!\Delta_1-\nu&=-\Delta_s,
g'=2gJ_1(\beta),
\quad \ \ \!
|01\rangle\leftrightarrow|10\rangle,
\\
\!\!\!\Delta_1-\alpha_2-\nu&=-\Delta_s,
g'=2\sqrt{2}gJ_1(\beta),
|02\rangle\leftrightarrow|11\rangle,
\end{align}
\end{subequations}
in which $|\Delta_s|\ll\nu$ is a small quantity, and $g'$ is the effective coupling strength in each interaction.

Ignoring the higher-order oscillating terms, then we further perform the unitary transformation with the transformation matrix $V(t)\!=\!\exp\{-i\Delta'tS_z/2\}$. Within the single-excitation subspace $\{|01\rangle, |10\rangle\}$ ($S_z=|01\rangle\langle01|-|10\rangle\langle10|$) or the two-excitation subspace $\{|02\rangle, |11\rangle\}$ ($S_z=|02\rangle\langle02|-|11\rangle\langle11|$), an effective two-level interaction Hamiltonian as same as Eq. (\ref{genH}) can both be obtained, and the correspondences of Hamiltonian parameters are
\begin{eqnarray} \label{corr12}
\Omega(t)\rightarrow g', \quad
\Delta(t)\rightarrow \Delta', \quad
\phi(t)\rightarrow \varphi'
\end{eqnarray}
where the effective phase $\varphi'=(\Delta_s-\Delta')t+\varphi-\pi/2$. It is thus expect to set the Hamiltonian parameters under different entangling interactions by following Eq. (\ref{OmegaPhi}) to execute the $5$-segment geometric trajectory correction. Therefore, we can also utilize the corrected $5$-segment geometric trajectory, that determined by the optimal regions of $\chi_1$ and $\chi_3$ with low error sensitivity, to implement our robust geometric $i\textrm{SWAP}^g$ and $\textrm{CZ}^g$ gates.

We next take the geometric $i\textrm{SWAP}^g$ gate as an illustration, which is enacted by bringing the entangling interactions between $|01\rangle$ and $|10\rangle$. First of all, through the gate parameters $\chi_0=\pi/2$ and $(\xi_0,\gamma_g)=(0,\pi/2)$ (or $(\xi_0,\gamma_g)=(\pi,3\pi/2)$) utilized to implement the $i\textrm{SWAP}^g$ gate, the optional ranges of $\chi_1$ and $\chi_3$ can be determined as $[0,\pi/2]$ and $[\pi/2,\pi]$ respectively. And then, we simulate the error sensitivity of our implemented $i\textrm{SWAP}^g$ gate under different $\chi_1$ and $\chi_3$ as shown in Fig. \ref{GateP2}(a), where the detuning error induced by the qubit-frequency drift, in the form of $\Delta'\rightarrow\Delta'+\delta'g'$, is considered as a typical error in superconducting quantum system. In Fig. \ref{GateP2}(b), it can be found that the optimal regions of $\chi_1=0.27\pi$ and $\chi_3=0.73\pi$ with low error sensitivity can be easily identified and used to implement geometric $i\textrm{SWAP}^g$ gate, which is obviously superior to the conventional $i\textrm{SWAP}^c$ gate in error suppression. In the same way, our geometric $\textrm{CZ}^g$ gate implemented based on the $5$-segment geometric trajectory correction (with $\chi_1\!=\!0$ and $\chi_3\!=\!0.9\pi$) can also demonstrate better fault-tolerant feature than the conventional $\textrm{CZ}^c$ gate as shown in Fig. \ref{GateP2}(c).
The above conventional $i\textrm{SWAP}^c$ and $\textrm{CZ}^c$ gates are realized based on the construction strategy of conventional quantum gate in Section II, which also utilize the two-qubit entangling interaction (as an effective two-level interaction) in the single-excitation subspace $\{|01\rangle, |10\rangle\}$ and two-excitation subspace $\{|02\rangle, |11\rangle\}$, respectively.

After verifying the fault-tolerant features of our geometric two-qubit gates implemented based on the $5$-segment geometric trajectory correction, we continue to evaluate their gate fidelity in realistic superconducting implementation, in which the overall influences of the neglected higher-order oscillating terms and the system decoherence are taken into account. The higher-order oscillating terms are the remaining terms in Eq. (\ref{EqHITwo}) except for the effective Hamiltonian used to realize the target entangling interactions, which will cause the unwanted perturbation coupling between $|01\rangle$ and $|10\rangle$ and leakage error from the computational basis $|11\rangle$ to the non-computational subspace $\{|02\rangle, |20\rangle\}$ \cite{TunCoupleExp2}, thus affecting the gate fidelity. In the subsequent simulation of the master equation, we set the anharmonicity of the other transmon $Q_2$ to $\alpha_2=2\pi\times280$ MHz and let it hold the same coherence time as the first transmon for simplicity, and the coupling strength between the two transmons is $g=2\pi\times8$ MHz. From the simulation results in Figs. \ref{GateP2}(d) and \ref{GateP2}(e), by optimizing the tunable parameters $\nu$ and $\beta$, our geometric $i\textrm{SWAP}^g$ and $\textrm{CZ}^g$ gates can both achieve a high fidelity of more than 99.60\%, which far reaches the fidelity threshold \cite{Nature14} of the surface code approach. As a result, we have taken the superconducting system as an example to illustrate the feasibility of our geometric trajectory correction scheme in physical implementation, which can make quantum gate possess stronger fault-tolerant feature while maintaining high gate fidelity. Again, our geometric trajectory correction scheme can potentially be matched in other quantum systems as well, requiring only a simple effective two-level structure.

\section{Discussion and Conclusion}

\emph{The same suppression effect on other error sources.}
In our scheme, we have taken the detuning error (such as the qubit-frequency drift in the superconducting circuit) as a typical example to verify the absolute robustness advantages of arbitrary type of geometric gate implemented based on geometric trajectory correction over conventional quantum gates. Remarkably, for other types of error sources, such as driving amplitude error in the form of $(1+\epsilon)\Omega(t)$, $\emph{ZZ}$ crosstalk in the form of $\zeta \sigma^1_Z\otimes\sigma^2_Z$, etc., ($\epsilon$ and $\zeta$ are error quantities), we can also effectively suppress them by executing our geometric trajectory correction. As an extension, it is worth noting that as the number $n$ of trajectory segments increases, our multi-segment geometric trajectory correction will be able to effectively suppress more error sources simultaneously, or demonstrate better suppression effect for certain errors.

\emph{Compatible with more optimal control technologies.}
During the execution of geometric trajectory correction, we did not impose any restrictions on the pulse shape of $\Omega(t)$ in each time segment. In addition, the different choices of pulse shape will not transform the robustness advantage of our geometric gate over conventional quantum gate. Therefore, we can continue to utilize the generality of our geometric scheme in selecting pulse shapes to be compatible with more optimal control techniques, including DRAG correction \cite{DRAG1} (verified in the main text), composite pulse \cite{ComP1,NGQC+Com}, dynamical corrected gates \cite{DynC1}, smooth optimal control \cite{SmoothC1}, etc., so as to further achieve hybrid fault-tolerant protection of geometric gate.

\emph{Extended to multi-level or multi-qubit holonomic control.}
Based on multi-level/multi-qubit system Hamiltonian, we can also determine a set of high-dimensional orthogonal evolution states governed by it, where the correspondence between the state-evolution parameters and the Hamiltonian parameters is associated through the Schr\"{o}dinger equation. Therefore, we can engineer multi-segment geometric evolution in the same manner to ensure that there are enough available evolution parameters to execute effective geometric trajectory correction against systematic errors. In this way, some potential schemes based on robust multi-level or multi-qubit holonomic control, such as the construction of holonomic qutrit/qudit and encoded multi-qubit gates, and state transfer, etc., can be realized.

In conclusion, we have developed a universal robust geometric control scheme based on geometric trajectory correction, which can sufficiently enhance the robustness of arbitrary types of geometric gates and show their absolute advantages over conventional quantum gate in suppressing systematic error. Our scheme effectively makes up for the shortcomings of conventional quantum control in robustness, and ensures that arbitrary types of geometric gates can all possess strong fault-tolerant feature, and high gate fidelity in physical implementation.

\bigskip

\acknowledgments

This work was supported by the Key-Area Research and Development Program of Guangdong Province (Grant No. 2018B030326001), the National Natural Science Foundation of China (Grants No. 12275090 and No. 11905065), Guangdong Provincial Key Laboratory (Grant No. 2020B1212060066), and the Guangxi Science Foundation (Grant No. AD22035186).


\end{document}